\documentclass[preprint]{aastex}

\usepackage{geometry}
\usepackage{graphicx}
\usepackage{amssymb}
\shorttitle{Two High-Mass Cores in ORI}
\shortauthors{Ren et al.}

\begin{document}

\title{Massive Quiescent Cores in Orion: V. The Internal Structures, Physical and Chemical Properties of the Two Extremely Dense Cores}


\author{Zhiyuan Ren\altaffilmark{1,2}, Di Li\altaffilmark{1,2,3} and}

\author{N. Chapman\altaffilmark{3,4}}

\altaffiltext{1}{National Astronomical Observatories, Chinese Academy of
Science, Chaoyang District Datun Rd A20, Beijing, China; Email:
renzy@nao.cas.cn, dili@nao.cas.cn}
\altaffiltext{2}{Key Laboratory of Radio Astronomy, Chinese Academy of Science}

\altaffiltext{3}{Space Science Institute, Boulder, CO, USA}

\altaffiltext{4}{Center for Interdisciplinary Exploration and Research in
Astrophysics (CIERA), Department of Physics \& Astronomy, 2145 Sheridan Road,
Evanston, IL 60208, USA; Email: nchapman@u.northwestern.edu}


\begin{abstract}
We present a high-resolution $(\sim1\farcs5)$ observational study towards two massive dust-and-gas cores, ORI8nw\_2 and ORI2\_6 in Orion Molecular Cloud using the Combined Array for Research in Millimeter-wave Astronomy (CARMA). In each region the 3.2 mm continuum emission exhibits a dense and compact dust core at the center with 1 to 3 solar masses. The cores have number densities exceeding $10^9$ cm$^{-3}$, which are among the highest volume densities observed in star-forming cores. In both regions the ${\rm N_2H^+}$ shows clumpy structures which are spatially displaced from the densest gas. In OIR8nw\_2 in particular, the ${\rm N_2H^+}$ shows a noticeable filament structure with a central cavity shell. The calculation for the dynamical state shows that this core can be potentially supported by the magnetic field against its gravitational instability, but the fragmentation might still occur and produce the observed ${\rm N_2H^+}$ clumps if the gas density exceeds $5\times10^7$ cm$^{-3}$ and this value is available within the observed density range. And the extremely high density at the core center suggests the super-Jeans condition and possibility for further fragmentation. For the chemical properties, the ${\rm N_2H^+}$-to-${\rm HCO^+}$ abundance ratios show a difference with that observed in infrared dark clouds. A combined analysis with the other Orion cores and the chemical model suggests that the different abundance ratios can be explained by the low CO abundances in our cores. To further reveal the evolution of such dense cores requires higher resolution and sensitivity. 
\end{abstract}


\section{Introduction}
Orion A Molecular Cloud (Orion A hereafter) is the closest star forming site which is in the vicinity of an OB cluster. It mainly consists of a dense central gas complex Orion KL, which contains plenty of YSOs surrounding the central OB stars, and two major giant filaments including the northern branch OMC-1,2,3 and southern branch OMC-4,5 to L1641 \citep{johnstone99,shimajiri11}. This entire system stores ample resources for the star formation at different mass scales and evolutionary stages. Numerous observations have been performed to explore its physical properties, including the large scale gas distribution, individual young stellar objects (YSOs), and the stellar feedback from the O-B clusters. The densest molecular gas in Orion A is mainly distributed along the giant filament which has a length of 4 degrees (13 pc) from north to south \citep[][etc.]{bally87,johnstone99}. These studies also show that the molecular gas and the star formation therein are severely influenced by the radiation from the central O-B stars. In the vicinity of the central stars, the molecular gas is compressed by the radiation pressure and gas expansion from the H{\sc ii} region, thereby has a tendency to form high-mass cores \citep{ikeda07}. Further to the south of OMC-1, the gas temperature, turbulence, and the molecular core masses all decrease \citep{bally87,tatematsu93,tatematsu08,buckle13} which might lead to an increased forming rate of low-mass stars \citep{buckle13}.

To investigate the environmental influence to the star formation, observations were performed towards individual young star forming sites such as L1641N \citep{fukui86}. \citet{galfalk08} systematically surveyed the YSOs in this region. Bright ${\rm H_2}$ outflows were found emanating from a number of stellar objects \citep{reipurth98,galfalk07,galfalk08}. A closer inspection showed that the CO outflow actually has more complex quadruple structure \citep[][SW07 hereafter]{sw07}. \citet{nakamura12} further showed that the stellar emissions might mainly trace the population at later stages (Class-I and II) while the less evolved YSOs may still be embedded in the dust envelope thus remain undetected. These younger sources may be critical for revealing the initial conditions in the Orion star forming regions, and may help us to understand the difference in the forming conditions for the low- and high-mass stars.  

In order to better characterize the initial conditions of the star formation in Orion, and evaluate the possibility of forming high-mass stars, surveys with dust and dense gas tracers were performed towards regions with potential dense and quiescent gas and faint infrared emissions \citep{li03,li07,li13}. \citet[][Paper II hereafter]{li07} identified 51 dust cores from seven fields in Orion A region. There is a fraction of the sources shared in common with other two surveys of Orion YSOs. \citet{manoj13} observed 21 protostars in far-infrared and measured their temperatures. The sample covered five sources in Paper II, including ORI1\_8, ORI1\_13, ORI2\_6, ORI2\_7, and ORI8nw\_2. They were all measured to have $T_{\rm bol}<70$ K thus classified as Class-0 objects. \citet{megeath12} carried out another extensive survey of Orion YSOs based on the Spitzer/IRAC emissions and covered 16 sources in Paper II. For all the shared sources, the IR counterparts were identified as "P(rotostellar)", indicating evolutionary stages earlier than Class-1.

On the other hand, a large fraction of the sources in Paper II (20 objects) exceed the equivalent Bonnor-Elbert (B-E) mass limit, suggesting that the core mass can be barely supported by the thermal pressure alone. In this case, the cores may possibly undergo gravitational collapse, or otherwise be stabilized by the turbulence and/or magnetic pressure. These two possibilities would critically determine the subsequent star formation thus require further examination. Moreover, the NH$_3$ observations on arc-second scales revealed possible temperature and turbulence decline towards the molecular core center \citep{li03}, suggesting that the cores have very weak internal heating thus may approximate prestellar stage. It calls for further observations with higher resolution and appropriate molecular tracers to better reveal their star-forming conditions. 

In this work, we present the millimeter interferometry observations towards two cores ORI8nw\_2 and ORI2\_6 in the Paper-II sample. These two cores have the highest mass ratio to the B-E limit. The 350 $\micron$ continuum emission in ORI8nw\_2 coincides with several YSOs in L1641N \citep{galfalk08}, while ORI2\_6 only have one faint IR source (HOPS 11 in \citet{manoj13} and No. 1100 in \citet{megeath12}). The two cores were observed in 3.2 mm continuum and ${\rm N_2H^+}$ and ${\rm HCO^+}$ $(1-0)$ lines. The ${\rm N_2H^+}$ can trace the dense quiescent gas due to its relatively stable abundance in the cold pre-stellar environment where many other species are frozen onto the dust grains (depleted). The ${\rm HCO^+}$ is one of the depleted molecules in the pre-stellar environment and would increase in gas phase as being heated by the protostars. 

The molecular and dust continuum emissions can reveal three aspects of the stellar evolution: 1) the gas distribution near the YSOs previously unresolved by the single dish; 2) the dynamical properties of the cores, including the turbulence and potential core collapse and fragmentation; and 3) the molecular abundances which can be further compared with the chemical models and other molecular surveys. We introduce the observations and data reduction in Section 2 and present the reduced data in Section 3. In Section 4 to 6, we analyze the kinematical, dynamical and chemical properties in the cores and discuss the influence from the outflow and star forming activities. A summary is given in Section 7. 


\section{Observation and Data Reduction}
We used the Combined Array for Research in Millimeter-wave Astronomy (CARMA) to observe the ${\rm N_2H^+}$ and ${\rm HCO^+}$ $J=1-0$ lines in our two targets.  These data were taken in 2007 and 2008. During the observations the CARMA array consisted of 15 antennae, including nine 6.1-meter dishes and six 10.4-meter dishes. We used two different array configurations: B array (baselines 100-1000 m) and C array (baselines 30-350 m). More details of the observations are shown in Table 1. 

We observed several different quasars for phase, bandpass, and flux calibration.  For the two 2007 datasets, we used 0541-056 as the phase calibrator. The bandpass and flux calibrators were 3C273 (2007-02-10) and 0530+135 (2007-03-12). In all 2008 observations we used 0530+135 as the phase calibrator. 3C84 was used as both a flux and bandpass calibrator, except for 2008-11-29 data, where 0423-013 was used for bandpass and flux. And on 2008-11-23, 0423-013 was used for the bandpass. The substitution became necessary when 3C84 was above an elevation of 80$^{\circ}$. The quasars for the flux calibration are routinely observed. Based on the variation in those values, we estimate our overall flux accuracy to be $\sim10\%$.

We set up our correlator so that each sideband had three windows all centered at the same frequency. In each sideband we had one wide-band window (approximately 500 MHz in 15 channels), one narrow-band window (approximately 8 MHz in 63 channels), and one medium-band window (approximately 32 MHz in 63 channels). In 2007 we only observed ${\rm N_2H^+}$, while in 2008 a slightly different correlator setup was used so that ${\rm N_2H^+}$ and ${\rm HCO^+}$ could be simultaneously observed in the two sidebands. The medium-band window was used to help identify emission in both molecules. This was especially useful for ${\rm N_2H^+}$ which has seven hyperfine components spaced over a $\sim15$-km s$^{-1}$ range \citep{caselli95}.  The narrowband window covers the frequency range of all seven hyperfine components and with better spectral resolution (0.4 km s$^{-1}$). Therefore, we focused on the narrowband data to analyze the two molecular lines.

The data were calibrated using the MIRIAD software package\footnote{http://carma.astro.umd.edu/miriad/}. After visually inspecting the data and flagging intermittent electronic noise on certain baselines and antennas, we performed calibrations for the bandpass (frequency-dependent variation of the gains) and flux using the calibrators as shown above. After the calibrations, we excluded the channel containing ${\rm N_2H^+}$ and ${\rm HCO^+}$ emissions to obtain the 3.2 mm continuum data. Then the different observing tracks (obtained from B and C arrays respectively) for each source are combined together and inverted from the (u,v) plane to the image plane using the robust parameter of 0.5. This value provides a compromise between the sensitivity of natural weighting and the lower side-lobes of uniform weighting. The combined data has a shortest baseline of 40 k$\lambda$, corresponding to an angular size of $50''$. The spatial structures larger than this scale will not be recovered in our CARMA maps. The dirty map produced by inversion was deconvolved using the maximum entropy method (MEM). The final clean map was created by convolving the model
image produced from MEM with a synthesized Gaussian clean beam (using RESTORE command in MIRIAD). For ORI8nw\_2, the clean beam is $1\farcs67\times1\farcs48$ with a position angle of $-54^\circ$ and for ORI2\_6 it is $1\farcs38\times1\farcs10$ with an angle of $83.5^\circ$. The images were corrected for efficiency of the primary beam which has a FWHM of $60''$.

Figure 1 and 2 show the channel maps of ${\rm HCO^+}$ and ${\rm N_2H^+}$ $J=1-0$ lines in the two regions. In ORI8nw\_2, we displayed the velocity range that covers the $F_1F=01\rightarrow12$ component. The other two groups of Hyper-fine Components (HFCs) $F_1=2\rightarrow1$ and $1\rightarrow1$ have have overall similar morphologies but blended components and are not shown here. The ${\rm HCO^+}$ emission in ORI8nw\_2 shows emission at the systemic velocity and in addition, highly redshifted emission around $V_{\rm lsr}=11$ km s$^{-1}$. In ORI2\_6 (Figure 2) the ${\rm N_2H^+}$ is much weaker, thus the strongest $F_1=2\rightarrow1$ transition is adopted for the display. The ${\rm N_2H^+}$ is marginally detected at $V_{\rm lsr}=7.0$ to 7.5 km s$^{-1}$, while the ${\rm HCO^+}$ emission is bellow the detection limit. The rms noise level $\sigma$ is measured from the emission-free areas in the channel images which is $\sigma=0.02$ Jy beam$^{-1}$. In ORI8nw\_2, it is equal to $T_{\rm b}=1.1$ K for the both lines. And in ORI2\_6, it corresponds to $T_{\rm b}=1.8$ K and 1.5 K for the ${\rm N_2H^+}$ and ${\rm HCO^+}$ lines, respectively.  

We also obtained images at other wavebands, including (1) the ${\rm C^{18}O~(2-1)}$ lines from a molecular line survey towards the Orion cores using the Caltech Submillimeter Observatory (CSO) (Ren et al. in prep.); (2) The mid-infrared images at four Spitzer/IRAC bands (3.6, 4.5, 5.8 and 8 $\micron$) and far-infrared images at Spitzer/MIPS 24 and 70 $\micron$ bands; (3) The JCMT/SCUBA 850 $\micron$ continuum images. The JCMT and MIPS images are only used to measure the emission intensities and not presented in the paper. The Spitzer images were retrieved from the Spitzer Heritage Archive\footnote{http://sha.ipac.caltech.edu/applications/Spitzer/SHA/}. The JCMT data were downloaded from the JCMT science archive\footnote{The James Clerk
Maxwell Telescope is operated by the Joint Astronomy Centre on behalf of the
Science and Technology Facilities Council of the United Kingdom, the Netherlands
Organisation for Scientific Research, and the National Research Council of
Canada. http://www.cadc.hia.nrc.gc.ca/jcmt/}.

\section{Reduced data}
\subsection{dust continuum emission}
In Figure 3 and 4, we show the 3.2 mm dust continuum emissions in ORI8nw\_2 and ORI2\_6, with the CSO/SHARC-II 350 $\micron$ continuum emission (gray scale) and the three-band IRAC image (blue=3.6 $\micron$, green=4.5 $\micron$, and red=8 $\micron$) overlaid. The molecular emissions are also shown on the figures and are described in Section 3.2. In ORI8nw\_2, the 3.2 mm continuum emission exhibits a single compact core located within the 350 $\micron$ emission region (Figure 3a). In ORI2\_6, the 3.2mm continuum peak coincides with the 350 micron emission within the pointing accuracy of the CSO. We manually adjusted the coordinates of the 350 $\micron$ emission by $4\farcs5$ to the east so that its emission peak overlaps with the peak of the 3.2mm continuum. 

At the IRAC bands (Figure 3c), ORI8nw\_2 contains five bright IR sources which are located within or near the 350 $\micron$ emission region. We encircled them with dashed lines and labeled them following the previous denominations \citep{galfalk08}. The two most evolved YSOs \citep[No.120 and 117 in][]{galfalk08}, which are bright at the $I$, $J$, and $K_s$ bands, are barely detected at the IRAC bands. In comparison, the other stellar objects (123, 124 etc.) can be clearly seen at the IRAC bands. Among them No.116 is the closest to the 3.2 mm peak (zero offset in Figure 3c). In ORI2\_6 (Figure 4c), there is only one single isolated IR source associated with the 3.2 mm continuum core.

The temperature, core mass and other physical parameters of the dust cores can be estimated from the dust continuum emissions. The results are summarized in Table 2. We first estimated the dust temperature by fitting the Spectral Energy Distribution (SED). We measured the flux densities of the cores at the IRAC 24 $\micron$ and MIPS 70 $\micron$ bands. At those bands, both ORI8nw\_2 and ORI2\_6 exhibit a compact core coincident with the 350 $\micron$ emission. The $850~\micron$ flux densities are taken from \citet{johnstone06}. In addition, for ORI8nw\_2, we also took the continuum flux density at 1.3 mm (3.1 Jy) from SW07, and at 2.0 mm (0.272 Jy) from \citet{chen06}. We adopt a gray-body emission model to fit the SED \citep{hildebrand83},

\begin{equation} 
F_{\nu}=\frac{M_{\rm core}\kappa_{\nu}B_{\nu}(T_{\rm d})}{g D^2}, \>
\end{equation}  
wherein $F_{\nu}$ is the flux density at frequency $\nu$, $M_{\rm core}$ is the total gas-and-dust mass of the core, $g=100$ is commonly adopted gas-to-dust mass ratio, $B_{\nu}(T_{\rm d})$ is the Planck function at $T_{\rm d}$, the dust temperature, $D=415$ pc is the source distance \citep{menten07,sandstrom07}, and $\kappa_\nu$ is the dust opacity and is assumed to have a power-law shape, i.e. $\kappa_\nu=\kappa_{\rm 230 GHz}(\nu/{\rm 230 GHz})^{\beta}$, with the reference value of $\kappa_{\rm 230 GHz}=0.9$ cm$^2$ g$^{-1}$~\citep{ossenkopf94}. The best-fit SED curves are shown in Figure 6. In ORI8nw\_2, we found that the emission from 70 $\micron$ to 1.3 mm can be fitted with $T_{\rm d}=24\pm3$ K and $\beta=1.6$. 

At shorter wavelengths (IRAC and MIPS 24 $\micron$ bands) the flux densities have excess above the SED curve which should arise from hot gas component at vicinity of the stars. The hot component is less accurately constrained by the available data points, but based on its local maximum intensity at around 24 $\micron$, we can roughly estimate $T=120$ K from Wien's displacement law. The 2.0 mm and 3.2 mm continuum emissions are both bellow the SED curve due to the extended emission being filtered by the interferometers. In ORI2\_6, the temperature (cold component) is only constrained by 75 $\micron$ (MIPS), 350 $\micron$ (CSO), and 850 $\micron$ (SCUBA) data. In order to reduce the parameters, we fit the SED assuming a same $\beta$ as in ORI8nw\_2. The temperature fitting gives $T_{\rm d}=19\pm4$ K. In both ORI8nw\_2 and ORI2\_6, the SED is characterized by a major cold dust component of $\sim20$ K and a faint hot one of $\gtrsim100$ K. Such SED shape is suggestive of a Class-0 object \citep{lada87}. By fitting the entire SED, \citet{manoj13} estimate an average bolometric temperature of $T_{\rm bol}=63$ and 59 K for ORI8nw\_2 (HOPS 182 therein) and ORI2\_6 (HOPS 11), respectively. These temperatures also suggest Class-0 stage according to their proposed threshold of $T_{\rm bol}=70$ K. The bolometric luminosities of the two cores are derived through the integral $L=4\pi D^2\int F_{\nu} {\rm d}\nu$. All the derived parameters are shown in Table 2.

We also calculated the total gas mass of the 3.2 mm continuum core using Equation (1), and obtained $M_{\rm core}=2.8$ and 1.6 $M_\odot$ for ORI8nw\_2 and ORI2\_6 respectively. The CARMA 3.2 mm core should mainly trace the densest central region while the CSO 350 $\micron$ core largely represent the extended envelope which takes up a considerable fraction of the core mass and is largely filtered out in the CARMA observations.

The H$_2$ column density $N({\rm H_2})$ and number density $n({\rm H_2})$ are also estimated from the continuum emissions. $N({\rm H_2})$ is calculated from the intensity at the continuum peak, which represents an average within one beam area. To estimate $n({\rm H_2})$, we assumed the core to have a spherical shape so that the dust column length is equal to the mean diameter (measured from the 4$\sigma$ contour size deconvolved with beam size as shown in Table 2). The derived column and number densities are also presented in Table 2. For the both sources, the 3.2 mm core is comparable to the beam size, suggesting that the cores are marginally resolved. In this case, the derived $N({\rm H_2})$ and $n({\rm H_2})$ would represent a lower limit. Both ORI8nw\_2 and ORI2\_6 exhibit remarkably high $N({\rm H_2})$ and $n({\rm H_2})$ for the 3.2 mm continuum core. Such high densities are only observed in a few sources, including the Orion KL cores \citep{beuther04} and some protostellar-disk candidates \citep{lopez11}. In comparison, the 350 $\micron$ continuum cores have much lower column densities, suggesting that the gas distribution is very concentrated at the core center so that the extended envelope only has a minor contribution to $N({\rm H_2})$ in spite of its large mass. In the opposite case, i.e., if the density profile were relatively flat, a larger fraction of the column density would be resolved out by the interferometers so that the $N({\rm H_2})$ and $n({\rm H_2})$ values would be significantly lower than those for the 350 $\micron$ continuum core. 

To better reveal the dust-and-gas distributions, we reconstructed the $N({\rm H_2})$ maps from the 3.2 mm and 350 $\micron$ continuum emissions, as shown in Figure 7. In calculation, we first convolve the 3.2 mm emission with the CSO beam, as to mimic observing the central dense core with the CSO beamsize at 350 $\micron$. Then the $N({\rm H_2})$ map was estimated from the convolved 3.2 mm and 350 $\micron$ continuum images respectively. As a result, the 3.2 mm emission exhibits a peak $N({\rm H_2})$ only slightly lower than that for the 350 $\micron$ emission. This is consistent with the speculation that the $N({\rm H_2})$ at the center is mainly contributed by the central dense core while the extended envelope only has a minor contribution.

The morphology of the envelope can be better revealed by subtracting the dense gas component (calculated from the 3.2 mm emission) from the overall $N({\rm H_2})$ distribution (from the 350 $\micron$ emission). The residual $N({\rm H_2})$ distributions in ORI8nw\_2 and ORI2\_6 are shown in Figure 7b and 7d, respectively. After subtraction, the residual $N({\rm H_2})$ map exhibits a quite flat profile within a spatial range of 1 arcmin, with a median value of $0.3\times10^{23}$ cm$^{-2}$. The residual $N({\rm H_2})$ slightly increases towards the northeast and southwest of the center as due to the slight elongation of the 350 $\micron$ core in this direction. In ORI2\_6, after the subtracting the 3.2 mm continuum core, $N({\rm H_2})$ also shows a roughly uniform distribution which has a slight decrease at the center and is elongated from NW to SE. 

By averaging $N({\rm H_2})$ at each radius, we calculated the one-dimensional $N({\rm H_2})$ profile for the 350 $\micron$ core, the 3.2 mm core and the envelope (residual). The three components are presented in Figure 7e and 7f. From the center to outside, the residual $N({\rm H_2})$ (envelope component) shows a small variation scale of 0.1 to 0.2 $\times10^{23}$ cm$^{-2}$. The figure also shows that in ORI2\_6 the 3.2 mm continuum core takes up a lower fraction of $N({\rm H_2})$ than in ORI8nw\_2. Considering the fact that ORI8nw\_2 contains several more evolved YSOs, this may reflect the trend that molecular cores are becoming more centrally concentrated during the core evolution \citep{bt12}. This trend is to be further examined from the density profiles in more pre- and protostellar cores in the Orion Molecular Cloud.

\subsection{Molecular lines}
The velocity integrated ${\rm N_2H^+}$ and ${\rm HCO^+}$ emissions in the both cores are presented in Figure 3 and 4. In ORI8nw\_2, the ${\rm N_2H^+}$ emission exhibits a filament with a spatial scale of $\sim60''$ (0.13 pc) from northeast to southwest. And the filament is resolved into six major gas clumps which are labeled as C1 to C6 in Figure 3a. Besides the major filament, the ${\rm N_2H^+}$ image also shows other two clumps to the north and south of the center cavity shell (labeled as C7 and C8 respectively), and in addition, some weak, dispersed gas fragments in the northwest. The ${\rm HCO^+}$ $(1-0)$ emission (Figure 3b) has a smaller extent than the ${\rm N_2H^+}$. It has three major emission peaks including one at the continuum peak and the other two in the southwest. The ${\rm HCO^+}$ $(1-0)$ clumps are aligned roughly parallel to the ${\rm N_2H^+}$ filament but has an overall $\sim5''$-shift to the north. There are two ${\rm HCO^+}$ clumps coincident with the IR sources No.123 and 116 while the ${\rm N_2H^+}$ apparently decreases towards these two sources.
 
In ORI2\_6, the ${\rm N_2H^+}$ emission mainly shows a weak and irregularly shaped gas clump located on the southwest side of the 3.2 mm continuum core, while the ${\rm HCO^+}$ $(1-0)$ does not show emission feature above the noise level (0.04 Jy beam$^{-1}$ km s$^{-1}$, as measured from the channel images). At the IRAC bands, ORI2\_6 has only one faint point source associated with the 3.2 mm continuum and the ${\rm HCO^+}$ emission (Figure 4c). The less active star formation in ORI2\_6 than in ORI8nw\_2 is some what unexpected as seen from their larger-scale environments. Over several-arcmin scales, ORI2 region exhibits multiple dust clumps aligned in filamentary structures, while ORI8nw\_2 is almost isolated at the same scale and more distant from the OMC center (Paper II, Figure 3 and 5 therein). A possible explanation is that star formation in ORI8nw\_2 is largely triggered by the cloud-cloud collision \citep[][see also Figure 4]{nakamura12}. While in ORI2, although there are multiple cores in a small region, they may currently maintain a stable structure, and have little interactions with each other.   

In ORI8nw\_2, both the ${\rm N_2H^+}$ and ${\rm HCO^+}$ emission exhibit multiple clumps at different velocities as shown in Figure 1 and 3. To more clearly elucidate the gas morphology and compare with the previous results, we show the velocity-integrated ${\rm N_2H^+}$ emissions and labeled the major gas structures in Figure 5. In general, the ${\rm N_2H^+}$ emission shows a filamentary structure from northeast to southwest which is intercepted by a central cavity-and-shell structure. The 3.2 mm core is located at the northern edge of the cavity wall. And the quadruple outflow system (SW07, with the directions shown in arrows) is closely correlated with the ${\rm N_2H^+}$ emission region. One can see that the collimated northeast-southwest outflow is nearly parallel to the ${\rm N_2H^+}$ filament while the more diffused north-south outflow is propagating through the filament and further to the south. Figure 5b shows the picture of the cloud-cloud collision on larger spatial scale \citep{nakamura12}. The collisional interface is from northeast to southwest, thus roughly parallel to the ${\rm N_2H^+}$ filament and the NE-SW outflow. The molecular gas may therefore be shaped by a net effect from the outflow and the cloud-cloud collision. The kinematical properties are discussed in more detail in Section 4.1.

We calculated the ${\rm N_2H^+}$ column density based on optical depth derived from fitting the HFCs. We assume that for each HFC, the optical depth as a function of the radial velocity has a Gaussian distribution \citep{myers96}, that is

\begin{equation} 
\tau_i(V)=\tau_i \exp[-\frac{(V-V_{\rm sys})^2}{2\sigma^2}]
\end{equation}

And the total optical depth is $\tau(V)=\sum{\tau_i(V)}$. The brightness temperature of the line emission would be 

\begin{equation} 
T_{\rm b}(V)=f[J(T_{\rm ex})-J(T_{\rm bg})](1-e^{-\tau(V)})
\end{equation}
where $f$ is the filling factor (assumed to be 1.0), $\tau_\nu$ is the optical depth of the line, $T_{\rm bg}=2.73$ K is the cosmic microwave background (CMB) temperature, and $J(T)$ is the Planck-corrected brightness temperature:
\begin{equation} 
J(T)=\frac{h\nu}{k}\frac{1}{e^{h\nu/kT}-1}
\end{equation}
wherein $k$ is the Boltzmann constant. The Hyper-fine fitting to the observed spectra was performed using the CLASS program in the GILDAS software package. The best-fit spectra are shown in Figure 8. For each core, the observed spectra can be reasonably fitted by adjusting the input parameters $\tau$, $\sigma$ and $T_{\rm ex}$. In ORI8nw\_2, we obtained $\tau({\rm F_1F=23-12})=0.5$ and $T_{\rm ex}=15$ K. In ORI2\_6 we have $\tau({\rm F_1F=23-12})\sim0.2$ and $T_{\rm ex}=16$ K. The excitation temperatures are close to the dust temperatures from the SED fitting. The column density is calculated from $\tau_\nu$ using
\begin{equation} 
N_{\rm tot}=\frac{8\pi \nu_0^3}{c^3}\frac{Q}{A_{\rm ul}g_{\rm u}}\frac{\exp(E_{\rm u}/kT_{\rm ex})}{\exp(h\nu_0/kT_{\rm ex})-1} \int \tau {\rm d}V
\end{equation}
wherein $Q$ is the partition function at the temperature $T_{\rm ex}$, $A_{\rm ul}$ is the Einstein Coefficient, $g_{\rm u}$ is the degeneracy of the upper level. For both molecules, we calculate the column density at two cases, (1) the maximum column density at their emission peak, and (2) the average column density within one CSO beam size $(9'')$ at the continuum peak. In calculation two temperature limits are considered, including the values from the SED fitting in this paper (Figure 6), and the bolometric temperatures from \citet{manoj13}. With two different values in beam sizes and temperatures, we altogether have four column-density estimates for each species in each source, which are all presented in Table 3. The ${\rm HCO^+}~(1-0)$ spectra are presented in right panels in Figure 8 and its column density is calculated in the same way. In ORI2\_6, the ${\rm HCO^+}~(1-0)$ spectrum towards the 3.2 mm continuum core shows a plausible emission feature, but the intensity is close to the noise level measured on the channel images. The derived column densities are therefore suggested to be an upper limit.   

The ${\rm N_2H^+}$ clumps are all absent in the 3.2 mm continuum emissions. We estimate an upper limit for their masses based on the noise level of the 3.2 mm continuum (1 mJy beam$^{-1}$) and the typical extent (20 arcsec$^2$). As a result, the clumps are estimated to have $F_{\rm 3.2 mm}<33$ mJy and $M<2~M_\odot$. 

We also examined the CO emission in the two cores. The CO column density is calculated from the ${\rm C^{18}O~(2-1)}$ line (Ren et al. in prep.). The CO spectra are shown in the right panels in Figure 8. The CO column density is calculated also at two temperature limits assuming an abundance ratio of ${\rm [^{12}CO]/[^{13}CO]}=75$ \citep[average value in the OMC,][]{knap81} and ${\rm [^{13}CO]/[C^{18}O]}=5$, namely ${\rm [^{12}CO]/[C^{18}O]}=375$. The derived $N({\rm CO})$ and CO abundances in both regions are also shown in Table 3. The CO abundances are comparable to the measurement in other orion molecular cores \citep{wilson11}.  

\section{Influence of the star formation and dynamical activities onto the molecular gas}
In ORI8nw\_2, the spatial morphology and velocity distribution of the molecular gas results from a net effect of the star-forming activities and dynamic processes, in particular the impact from the outflow. The significance of these factors are discussed as following.

\subsection{The impact from the outflow}
To better reveal the potential influence from the outflow, in Figure 9 we plot the two outflow lobes in contours (SW07) together with the ${\rm N_2H^+}$ and ${\rm HCO^+}$ emissions (gray scale). For the ${\rm N_2H^+}$, the major filament structure is nearly parallel to the outflow axis, but with an overall offset to the south. The central cavity of the ${\rm N_2H^+}$ is filled by the southern branch of the redshifted CO outflow lobe. The outflow extends to the cavity wall and propagates further to the south, as indicated by the arrow in Figure 9a. The red lobe has a small fraction on its southern edge overlapped with the ${\rm N_2H^+}$, while the blue lobe propagates throughout the central cavity wall to the northeast, and has cleared out almost all the ${\rm N_2H^+}$ along its pathway. 

The different outflow components seem to have different chemical consequences. The more extended southern red lobe is sweeping the gas onto the cavity wall while the collimated NE-SW flow is likely causing more disruption of ${\rm N_2H^+}$. As shown in SW07 (and also seen in Figure 1), the collimated NE-SW flow has a velocity distribution from $V_{\rm lsr}=10$ to 20 km s$^{-1}$, while the southern branch of the red lobe becomes weak at $V_{\rm lsr}>12$ km s$^{-1}$. Although the actual outflow velocities are uncertain due to the inclination angle, the NE-SW flow is likely to have higher velocities due to the compact and collimated shape. This suggests that the outflow velocity should be a key factor to determine the chemical effect. 

Since the ${\rm N_2H^+}$ and ${\rm HCO^+}$ both appear to be affected by the outflow, their kinematic properties should be further examined. We plotted the intensity-weighted radial-velocity field (first moment map) and the velocity-dispersion (second moment map) of the ${\rm N_2H^+}$ in Figure 9b and 9c, respectively. The two maps are calculated from the single-peaked $F_1F=01-12$ component (f=93.17613 GHz). As shown in Figure 9b, the velocity field of ${\rm N_2H^+}$ shows no significant deviation from the systemic $V_{\rm lsr}$, nor any velocity gradient over the emission region. It only shows moderate redshift features at several positions overlapped with the outflow, including the southern edge of the cavity wall, and the northwestern edge of the filament where the red lobe (dashed contours) goes over. 

\citet{nakamura12} identified three concentrically expanding shells on larger scales which are also plotted in Figure 5b. A question is whether the cavity wall in the ${\rm N_2H^+}$ emission represent another recently formed shell in a similar physical process. To our expectation, an expanding shell would exhibit both blue- and redshift feature as well as an increased velocity dispersion since it is moving towards all directions. However, the cavity wall only shows redshifted emission (Figure 9b) and a slightly higher velocity dispersion (up to 1 km s$^{-1}$, Figure 9c). Inside the cavity wall, the ${\rm N_2H^+}$ is not detected above the uncertainty level of 0.06 Jy beam$^{-1}$ km s$^{-1}$, which limits the column density to be $N({\rm N_2H^+})<4.5\times10^{12}$ cm$^{-2}$. This upper limit is almost two orders of magnitudes smaller than $N({\rm N_2H^+})$ at the cavity wall, suggesting that the cavity wall is more likely a 2-dimensional ring-like structure rather than a spherical shell. 

The ${\rm HCO^+}$ is closely associated with the outflow for its velocity distribution. Its spatial correlation with the outflow, radial velocity, and velocity dispersion map are shown in Figure 9d to 9f. Figure 9d shows that the three major ${\rm HCO^+}$ gas clumps reasonably coincide with the local emission peaks in the red lobe. Figure 9e shows that the redshift pattern in ${\rm HCO^+}$ well coincides with the spatial extent of the CO outflow. The fraction of the ${\rm HCO^+}$ covered by the outflow also shows increased velocity dispersion (increasing from the average value of $\sigma=2$ km s$^{-1}$ to  $\sigma=3$ km s$^{-1}$, Figure 9f). From Figure 1 we see that a bulk of the redshifted ${\rm HCO^+}$ emission appears at $V_{\rm lsr}=10$ to 12 km s$^{-1}$ and becomes weaker at lower velocities (8.5 to 9.5 km s$^{-1}$). This suggests that the ${\rm HCO^+}$ might be largely produced in the (redshifted) outflow rather than merely entrained from the ambient gas.

\subsection{The emission from the stellar objects}   
Both ORI8nw\_2 and ORI2\_6 have considerable total masses for their CSO 350 $\micron$ continuum cores (Table 2). However, the ${\rm N_2H^+}$ emissions reveal low-mass gas clumps at smaller scales, and the 3.2 mm continuum core also has gas masses of only $\sim1 M_\odot$, suggesting that the two cores are currently only forming low-mass stars. Although the CARMA observations may have largely missed extended structures, their low bolometric luminosities (Table 2) also suggest the absence of high-mass stars.


The IR sources in ORI8nw\_2 are found to have a total luminosity even lower than the dust core luminosity. There are mainly 6 IR point sources observed at the IRAC bands labeled as No.115, 116, 117, 120, 123, 124 \citep[][also see Figure 3c]{galfalk08}. We can get their SEDs by interpolating the measured flux at four IRAC bands \citep[][Table 11 therein]{galfalk08}, and then derive the integrated luminosities. As a result, the IR sources all have luminosities at $10^{-1}~L_{\odot}$ scale. Among the IR sources, No. 117 and 120 are very faint at IRAC bands but are detected in the $I$, $J$, and $K_s$ bands and estimated to have 0.15 and 0.02 $L_{\odot}$ (Table 6 therein). For ORI2\_6, the central IRAC source is measured to have 0.12 $L_{\odot}$, thus also much fainter than the entire dust core ($20~L_{\odot}$). On the other hand, based on the stellar reddening measured by \citet{galfalk08}, we obtained a moderate extinction of $A_{[4.5~\micron]}=1.1$ for ORI8 region. This suggests that the stellar objects with high luminosities are not likely to be largely obscured at the IRAC bands. Considering the faintness of the IR sources, the bolometric luminosities of the dust cores should be mainly contributed by the external heating.

\subsection{The cloud-cloud collision}
Another potential influence to the molecular distribution comes from the cloud-cloud interaction \citep{nakamura12}, which might have compressed the ${\rm N_2H^+}$ gas, causing it to be extended nearly in the same orientation with the collisional interface(Figure 7b). The interaction might also aid the mass accumulation, causing ORI8nw\_2 to become the most massive core over one-square-degree region. But the interaction did not lead to high-mass star formation in this region. 


\section{Dynamical conditions}
Paper II investigated the gravitational instabilities of ORI8nw\_2 and ORI2\_6 based on the assumption that the molecular cores are gravitationally bound. Using the currently observed molecular lines, we can better estimate the dynamical state in two major aspects: (1) whether the cores are gravitationally bound (the state of virial equilibrium); (2) the possibility for gravitational collapse and fragmentation. 
The virial mass can be estimated from 
\begin{equation} 
M_{\rm vir}=\frac{5}{\alpha\beta}\frac{\sigma^2 r}{\rm G},
\end{equation}
wherein $r$ is the average radius. $\sigma$ is the velocity dispersion. $\beta=\arcsin e/e$ is the geometry factor determined by eccentricity $e$. $\alpha=(1-k_\rho/3)/(1-2k_\rho/5)$ for a power-law density profile $\rho\propto r^{-k_\rho}$ \citep{bertoldi92,mckee92}, and ${\rm G}$ is the constant of gravity. We adopt $k_\rho=2$ which characterizes a static singular isothermal sphere \citep{shu77}, and $e=0$ since the both the 350 $\micron$ and the 3.2 mm cores are roughly spherical. And we considered the velocity dispersion from both the ${\rm N_2H^+}$ and ${\rm C^{18}O}$ lines for the calculation, which would reflect the turbulence in the quiescent dense gas and the more extended gas components, respectively. The estimated virial masses are shown in Table 2. 

In ORI8nw\_2, from the velocity dispersion measured from the ${\rm N_2H^+}$ line, we can get $M_{\rm vir}=8~M_\odot$ for the 350 $\micron$ core, which is largely smaller than the LTE gas mass. Using the ${\rm C^{18}O}$ line width instead, the derived $M_{\rm vir}~(20~M_\odot)$ is still lower than $M_{\rm core}$, suggesting that the core should be gravitationally bound. One uncertainty is that the calculation of $M_{\rm core}$ depends on the adopted temperature. For example, using $T_{\rm dust}=63$ K \citep{manoj13}, the core mass would decrease to $3~M_\odot$ which is significantly smaller than $M_{\rm vir}$. For ORI2\_6, using $T_{\rm d}=59$ K \citep{manoj13} would provide $M_{\rm core,350\micron}=1.0~M_\odot$ which is also smaller than $M_{\rm vir}$. 

The dense pre- and protostellar cores should be close to virial equilibrium. For example, \citet{ikeda07} studied a large sample of ${\rm HCO^+}$ cores over the entire Orion A filament and found them to be mostly virialized. The ${\rm HCO^+}$ cores on their observational scales are measured to have densities of only $10^3$ to $10^4$ cm$^{-2}$. In comparison, the compact massive dense cores on smaller scales should be more gravitationally bound. The suggestion is that the major fraction of the 350 $\micron$ continuum core should have moderately low temperatures, likely in a range of 20-30 K, so that the derived core masses can exceed the virial masses. In particular, assuming an ideal virial equilibrium, i.e., $M_{\rm core,350\micron}=M_{\rm vir}$, we can obtain $T_{\rm d}\sim25$ K from Equation (1). \citet{miettinen12} also used two-temperature model to fit the SED of several Class-0 Orion cores, and obtained $T\sim10$ K for the cold components which are consistent with the temperature estimate from the molecular lines. The high temperature values for our two cores should still be valid, but mainly reflect the hot gas components surrounding the stars, and can be distinguished from the cold gas in the SED fitting (Figure 5).

We also examined the virial state of the central 3.2 mm continuum cores. For ORI8nw\_2, since the ${\rm HCO^+}$ emission is closely associated with the 3.2 mm continuum cores, we adopt its line width and the upper limit of $r=1.3''$ and then estimated $M_{\rm vir}\leq0.6~M_\odot$ which is even smaller than $M_{\rm core,3.2mm}$ estimated at $T=63$ K. Thus the 3.2 mm continuum core may also be gravitationally bound unless its temperature is even higher so that the gas mass is overestimated. In ORI2\_6, the velocity dispersion for the 3.2 mm continuum core is less well determined due to the absence of molecular lines at the core center. Using the ${\rm N_2H^+}$ line detected near the 3.2 mm continuum core, we can get $M_{\rm vir}=0.4~M_\odot$. In comparison, it has a total gas mass of 0.3-0.8 $M_\odot$, thus should be close to the virial equilibrium.

The second aspect of the dynamical state is the stability against collapse and fragmentation. The critical mass $M_{\rm c}$ to be sustained by the internal force is $M_c=M_J+M_\Phi$, wherein $M_J$ and $M_\Phi$ represent the mass to be supported by the random gas motion (thermal motion and turbulence) and magnetic pressure, respectively. The two components can be estimated using the same procedures in \citet[][and the references therein]{li13}. The Jeans mass for a non-magnetic isothermal cloud \citep{bonnor56,mckee92} is 
\begin{equation}
M_J=1.182 \frac{\sigma^4}{G^{3/2}P_{\rm ic}^{1/2}}
\end{equation}
wherein $P_{\rm ic}$ is the external pressure, and can be estimated as
\begin{equation}
P_{\rm ic}=n_{\rm ic}\mu m_{\rm H} \sigma_{\rm ic}^2
\end{equation}
wherein $m_{\rm H}$ is the atomic hydrogen mass and $\mu=2.33$ is the average molecular weight \citep{myers83}. In calculation we adopt an environmental velocity dispersion of $\sigma_{\rm ic}=1$ km s$^{-1}$, and density of $n_{\rm ic}=0.8\times10^4$ cm$^{-3}$ as measured from the extended ${\rm H^{13}CO^+~(1-0)}$ emission \citep{ikeda07}. The external pressure is calculated to be $P_{\rm ic}=2.3\times10^7$ K cm$^{-3}$. Assuming that the observed ${\rm N_2H^+}$ line width represents the total random gas motion (thermal motion and turbulence), we can get $\sigma_{\rm tot}=0.53$ km s$^{-1}$ and $M_{\rm J,tot}\sim4~M_\odot$. The total Jeans mass is much lower than the LTE gas mass, suggesting that the random gas motion (thermal motion and turbulence) can hardly support the core against collapse. 


The maximum mass to be stabilized by the $B$ field is 
\begin{equation}
M_{\rm \Phi}=c_{\rm \Phi}\frac{\pi B r^2}{G^{1/2}}
\end{equation}
where $c_{\rm \Phi}\sim 0.12$ \citep{tomisaka1988}. \citet{crutcher99} measured $B=0.36$ mG in the OMC-1 region where the gas density is $10^{5.9}$ cm$^{-3}$, and later derived a power law of $B\propto n^{0.65}$ from a number of Orion cores \citep{crutcher10}. Using this result, we can estimate $B=0.66$ mG for gas density in ORI8nw\_2 ($2.1\times10^6$ cm$^{-3}$). \citet{norris84} observed OH masers in Orion-KL and derived $B\sim 3$ mG. \citet{tang10} observed the polarized dust continuum emission also in Orion-KL and suggested a field intensity of $B\geqslant3$ mG. For ORI8nw\_2, we can then derive $M_{\rm \Phi}=4.4$ to $20~M_\odot$ as the $B$-field varies from 0.66 to 3 mG. This suggests that the $B$-field is possible to provide a considerable support to the core if the field strength is at the milli-Gauss level.

If the collapse is ongoing, the typical fragmentation scale, namely the Jeans length can be estimated and compared with the ${\rm N_2H^+}$ clumps to see whether they are close to each other. The Jeans length is calculated using $\lambda_J=\sqrt{\pi \sigma^2/G \rho}$ \citep{jeans02}, where $\rho=\mu m_0 n_{\rm H_2}$ is the gas density. Using the gas density of the 350 $\micron$ core and $\sigma$ from the ${\rm N_2H^+}$ line width, the Jeans length is estimated to be $\sim10$ arcsec. For the magnetic pressure, assuming a velocity dispersion comparable to the Alfv{\'e}n velocity $v_A=B/\sqrt{4 \pi \rho}$, we can get its equivalent Jeans length of $\lambda_{J,B}=15''$ to $100''$ as $B$ field varies from 0.66 to 3 mG. We see that as $B$ field increasing ($>1$ mG), the total Jeans length would be dominated by the magnetic component and largely exceeds the ${\rm N_2H^+}$ clump scales ($5''$ to $10''$), thereby stabilize the core against fragmentation. Yet the fragmentation may still occur based on the evidence from three aspects. 

First, the Jeans length would become lower as the gas density increases towards the core center, while the $B$-field only moderately increase with the density if it follows the power law in \citet{crutcher10}. Using the gas density in the central dense core ($10^9$ cm$^{-3}$), we can get $\lambda_{J,B} \sim1.0''$. To reach a fragmentation scale rightly at ${\rm N_2H^+}$ clump size ($\sim5''$), a density of $\sim5\times10^7$ cm$^{-3}$ is required which is achievable as comparing to the overall density range. 

Second, the magnetic turbulence can in the mean time cause density fluctuation thus motivate the fragmentation \citep[][etc.]{hanawa93,takahashi13}. Using the method in \citet[][Equation (5) to (7) therein]{takahashi13}, we can obtain a fragmentation length of $\lambda_{\rm frag}=40''$ to $1.0''$ as the gas density varying from $10^6$ to $10^9$ cm$^{-3}$. This spatial range is compatible with the ${\rm N_2H^+}$-clump scales. In fact, multiple clumps with sizes around $r\sim5''$ have already been observed in Orion KL \citep{tang10}. And the clumps further exhibit substructures at even smaller scales. The condensations are apparently smaller than $\lambda_{J,B}$ derived from the $B$ field strength therein, suggesting that the $B$-field might not effectively halt the fragmentation or might even provide an enhancement.

Third, if there is currently no fragmentation in ORI8nw\_2, the observed ${\rm N_2H^+}$ clumps may have to be solely shaped by the outflow or/and by the cloud-cloud interaction. This case is not fully consistent with the fact that the ${\rm N_2H^+}$ clumps have small velocity gradient and dispersion, and at the mean time spatially displaced from the observed outflow (Figure 9a). 


The possibility of the fragmentation can be further examined by looking at internal structures in other similar Orion cores with sensitive dense gas tracers, and more accurately measuring the $B$-field strength and its spatial correlation with the gas structures.

\section{The chemical evolution with ${\rm N_2H^+}$ and ${\rm HCO^+}$}


In this section we discuss the chemical properties related with ${\rm N_2H^+}$ and ${\rm HCO^+}$. In most cases, the ${\rm N_2H^+}$ would closely trace the the dense quiescent gas \citep[e.g.][]{pirogov03,sanhueza12}. But the outflow and stellar heating may cause a deviation from the general trend, as observed in our two cores. As shown in Figure 3, in ORI8nw\_2, the ${\rm N_2H^+}$ is spatially displaced from the 3.2 mm continuum and the ${\rm HCO^+}$ emissions. In ORI2\_6, the weakly detected ${\rm N_2H^+}$ emission is also offset from the 3.2 mm continuum core (Figure 4). 

In ORI8nw\_2, the ${\rm HCO^+}$ mainly traces the outflow and lies in the close proximity of the ${\rm N_2H^+}$. Towards the stellar objects No.123 and 116 (Figure 3b), the ${\rm HCO^+}$ exhibits a local emission peak while the ${\rm N_2H^+}$ obviously declines. Therefore the ${\rm N_2H^+}$ is likely being directly converted to ${\rm HCO^+}$ via the reaction 
\begin{displaymath}
{\rm N_2H^++CO\rightarrow HCO^++N_2~~~~(a)}
\end{displaymath}
The reaction is endothermic, thus when temperature is low, the energy source like outflow would sensitively reduce the ${\rm N_2H^+}$ and increase the ${\rm HCO^+}$, causing the observed features. Besides reaction (a), the ${\rm HCO^+}$ can also be produced from 
\begin{displaymath}
{\rm CO+H_3^+\rightarrow HCO^++H_2~~~~(b)}
\end{displaymath}
Reaction (a) and (b) are both taking place in the IRDCs \citep{vasyunina12}. But (b) might be less important in producing the ${\rm HCO^+}$ in our dense cores since the ${\rm HCO^+}$ shows a close dependence on the ${\rm N_2H^+}$ distribution. In less dense and more extended gas component, reaction (b) would be more significant as the CO becomes depleted and ${\rm H_3^+}$ also becomes more abundant due to the photo dissociation. This speculation is consistent with the single-dish observation \citep{ikeda07} which shows broad ${\rm H^{13}CO^+}$ emission over ORI8 region (L1641N) with a spatial extent of 5 arcmin. The gas density at this scale was measured to be only $10^3$ to $10^4$ cm$^{-3}$, which is largely different from the gas component in the compact dense cores.

The abundance ratio between ${\rm N_2H^+}$ and ${\rm HCO^+}$ (${\rm [N_2H^+]/[HCO^+]}$) can reflect the chemical evolution in the molecular gas. \citet{sanhueza12} measured ${\rm [N_2H^+]/[HCO^+]}$ in 37 Infrared Dark Clouds (IRDCs) and obtained a distribution from 0.07 to 0.12. ${\rm [N_2H^+]/[HCO^+]}$ in the IRDCs exhibit a plausible variation among the different stages, but the entire variation range is lower than ${\rm [N_2H^+]/[HCO^+]}$ in our two cores. In order to explain this difference, we referred to the chemical model presented by \citet{jorgensen04}. The model estimated the evolution of ${\rm [N_2H^+]}$ and ${\rm [HCO^+]}$ as a function of the CO abundance (Figure 16 therein) in a physical condition of $n({\rm H_2})=10^6$ cm$^{-3}$ and $T=20$ K which is comparable to the conditions in our cores. The major qualitative prediction of the model is that an increased [CO] would continuously decrease ${\rm [N_2H^+]}$ while increase ${\rm [HCO^+]}$. This trend can be compared with the observed abundances.

In Figure 10, we plot ${\rm [N_2H^+]/[HCO^+]}$ as a function of [CO] derived from the model, and overlaid the observed values in our two cores. The arrow on the ORI2\_6 data represents lower limit of ${\rm [N_2H^+]/[HCO^+]}$ due to the barely detected ${\rm HCO^+}$ emission. In the Figure, the abundances calculated at high temperatures \citep{manoj13} are also presented in open squares. At both temperature limits, the observed data are consistent with general trend that ${\rm [N_2H^+]/[HCO^+]}$ declines with [CO]. This proves that the CO should be a major controller for the ${\rm HCO^+}$ production. At low temperatures, the derived abundance ratios are close to the modeled curve, while at high temperatures, ${\rm [N_2H^+]/[HCO^+]}$ become much higher. This is within expectation since the model parameters are also originated from the observed values at low temperatures. There are three IRDCs in \citet{sanhueza12} also with available [CO] measurement \citep{vasyunina11}.  Their ${\rm [N_2H^+]/[HCO^+]}$ and [CO] are presented together in Figure 10. The IRDCs have [CO]$\gtrsim10^{-4}$ and ${\rm [N_2H^+]/[HCO^+]}\gtrsim0.1$ thus are close to the modeled curve. This suggests that the different ${\rm [N_2H^+]/[HCO^+]}$ ratios between our cores and the IRDCs can be explained by their different CO abundances. 

In addition, we also searched for other Orion cores with available data. As a result, two cores in BN-KL region \citep{ungerechts97} and another one in Orion B \citep{miettinen12} were found. The three additional Orion cores also show relatively high [CO] and small ${\rm [N_2H^+]/[HCO^+]}$ compared to ORI8nw\_2 and ORI2\_6. However, the Orion cores together show a more rapid decline trend than the model prediction, i.e., at [CO]$=10^{-4}$, the modeled ${\rm [N_2H^+]/[HCO^+]}$ approaches 0.1, while the three additional Orion cores have reached ${\rm [N_2H^+]/[HCO^+]}\sim10^{-2}$. From examining the two species individually, we see that the difference is mainly due to ${\rm [HCO^+]}$ which largely exceeds the model prediction, as well as the IRDC values. The ${\rm HCO^+}$ enhancement in these Orion cores can be tentatively attributed to the heating from outflow/shock and UV radiation which all broadly exist in Orion. It still calls for more molecular lines and data from more Orion cores to examine the related chemistry, in particular whether ${\rm [N_2H^+]/[HCO^+]}$ is also affected by other factors besides the CO abundance.
 
\section{Summary and Conclusions}
Using the CARMA, we studied the two Orion molecular cores ORI8nw\_2 and ORI2\_6 the masses of which exceed the thermally stable B-E limit. We observed them in ${\rm N_2H^+}~(1-0)$ and ${\rm HCO^+}~(1-0)$ and examined their molecular distribution, chemistry and gas kinematics. Our main findings are:

(1) As shown in the 3.2 mm continuum emission, for each core, the central region is extremely dense, with ${\rm n(H_2)>10^9~cm^{-3}}$ and has a compact size of several $10^2$ AU. The central dense core is likely to have a super-Jeans state and undergo further fragmentation. The dense core is surrounded by a flat envelope which extends to 4000-7000 AU and has a relatively lower density ($\sim10^6$ cm$^{-2}$). 

(2) In both regions, the ${\rm N_2H^+}$ is spatially displaced from the dense 3.2 mm continuum core. This should be mainly due to the disruption of ${\rm N_2H^+}$ during the chemical evolution. In ORI8nw in particular, the strong outflow has cleared most of the ${\rm N_2H^+}$ along its pathway, in the mean time generated large amount of ${\rm HCO^+}$. 

(3) ${\rm N_2H^+~(1-0)}$ emission in ORI8nw\_2 is resolved into multiple clumps which are aligned in a 50-arcsec filamentary structure with a central cavity. The outflow, cloud-cloud collision, fragmentation could all be involved in forming these gas clumps. Since the outflow impact or the cloud-cloud collision is not evidently shown in the velocity field of the ${\rm N_2H^+}~(1-0)$, the fragmentation might be more important in producing the clumps. The analysis for the Jean instability shows that the magnetic pressure can considerably stabilize the core. But the fragmentation is still possible if the gas density is sufficiently high ($n({\rm H_2})\sim5\times10^7$ cm$^{-3}$). ORI8nw\_2 presents an example that a high-mass core is only forming low-mass stars possibly due to the fragmentation.

(4) The molecular abundances in the Orion cores are roughly consistent with the modeled trend that the ${\rm [N_2H^+]/[HCO^+]}$ ratio declines with [CO]. But together with other Orion core data, the decline of ${\rm [N_2H^+]/[HCO^+]}$ shows a more significant decline than that the modeled trend in quiescent gas. This may be due to the stellar emission and the outflow in OMC which increases the production of ${\rm HCO^+}$.

The high resolution allows us to reveal the structures and chemical evolutions of two massive Orion cores. The effects of cloud evolution and protostellar feedback are revealed in details. There are still some key properties to be investigated, in particular whether the central dense core would undergo fragmentation or monolithic core collapse. Considering their ample star-forming activities and nearby distance, these Orion cores will be ideal targets for future ALMA observations.

\section*{Acknowledgment}
We are grateful to Dr. F. Nakamura and Dr. T. Stanke for their kind and valuable advice and support. We would like to thank the anonymous referee for the careful inspection of the manuscript and constructive comments. We also thank the staff who maintain the MIRIAD MacPort package. This work is supported by the China Ministry of Science and Technology under State Key Development Program for Basic Research (973 program) No. 2012CB821802, the National Natural Science Foundation of China No. 11373038, No. 11373045, Strategic Priority Research Program "The Emergence of Cosmological Structures" of the Chinese Academy of Sciences, Grant No. XDB09010302, and the Young Researcher Grant of National Astronomical Observatories, Chinese Academy of Sciences.

\begin{table}
{\small
\centering
\begin{minipage}{160mm}
\caption{Observational information.}
\begin{tabular}{lllcccl}
\hline\hline
Source$^a$     &  RA                       &  Dec$^b$                  &  Date Observed     &  Array      &  Time$^c$   &  Molecular   \\
\quad          &  (J2000)                  &  (J2000)                  &  (YYYY-MM-DD)      &  Config.    &  (hours)    &  Transition  \\
\hline
ORI2\_6        &  ${\rm 05^h35^m13\fs13}$  &  ${\rm -05\arcdeg57\arcmin58\farcs5}$  &  2007-02-10        &  C          &  4.9        &  ${\rm N_2H^+}$  \\
\quad          &  \quad                    &  \quad                    &  2007-03-12        &  C          &  1.7        &  ${\rm N_2H^+}$  \\
\quad          &  \quad                    &  \quad                    &  2008-11-24        &  B          &  3.0        &  ${\rm N_2H^+}$, ${\rm HCO^+}$  \\
\quad          &  \quad                    &  \quad                    &  2008-11-29        &  B          &  3.8        &  ${\rm N_2H^+}$, ${\rm HCO^+}$  \\
\quad          &  \quad                    &  \quad                    &  2008-12-09        &  B          &  2.6        &  ${\rm N_2H^+}$, ${\rm HCO^+}$  \\
ORI8nw\_2      &  ${\rm 05^h36^m18\fs6}$   &  ${\rm -06\arcdeg22\arcmin10\farcs4}$  &  2008-11-05        &  C          &  3.4        &  ${\rm N_2H^+}$, ${\rm HCO^+}$  \\
\quad          &  \quad                    &  \quad                    &  2008-11-11        &  C          &  2.3        &  ${\rm N_2H^+}$, ${\rm HCO^+}$  \\
\quad          &  \quad                    &  \quad                    &  2008-11-23        &  B          &  5.4        &  ${\rm N_2H^+}$, ${\rm HCO^+}$  \\
\quad          &  \quad                    &  \quad                    &  2008-12-18        &  B          &  4.0        &  ${\rm N_2H^+}$, ${\rm HCO^+}$  \\
\hline    
\end{tabular}\\
$a.${ Core names follow that in \citet{li07} (Paper II).} \\
$b.${ Positions are the phase tracking centers.} \\
$c.${ The time indicates the on-source integration time.} \\
\end{minipage}
}
\end{table}

\begin{table}
{\scriptsize
\centering
\begin{minipage}{140mm}
\caption{Physical properties of the two dust-and-gas cores.}
\begin{tabular}{lccc}
\hline\hline
Parameters                          &  ORI8nw\_2                         &  ORI2\_6                           &  Unit         \\
\hline                                                                                                                        
Peak Position                       &  ${\rm 05^h36^m18\fs79}$           &  ${\rm 05^h35^m13\fs43}$           &  \quad        \\ 
\quad                               &  ${\rm -06\arcdeg22\arcmin10\farcs40}$  &  ${\rm -05\arcdeg57\arcmin57\farcs9}$           &  \quad        \\
$T_{\rm dust}$                      &  $23\pm3$                          &  $19\pm4$                          &  K            \\
$T_{\rm ex}({\rm N_2H^+})$          &  9                                 &  16                                &  K            \\
$L_{\rm bol}$                       &  60                                &  20                                &  $L_{\odot}$  \\
$L_{\rm stars}^a$                   &  0.17                              &  0.13                              &  $L_{\odot}$  \\
$M_{\rm core,350\micron}^b$         &  46                                &  12                                &  $M_{\odot}$  \\
$M_{\rm core,3.2mm}$                &  2.8                               &  1.6                               &  $M_{\odot}$  \\
$R_{\rm core,350\micron}$           &  $15''$ (6460 AU)                  &  $11''$ (4600 AU)                  &  $-$          \\
$R_{\rm core,3.2mm}$                &  $\lesssim1.3''$ (520 AU)          &  $\lesssim1.2''$ (490 AU)          &  $-$          \\
$M_{\rm vir,350\micron}^c$          &  8.4,34                            &  2.8,7.0                           &  $M_{\odot}$  \\
$M_{\rm vir,3.2mm}^c$               &  0.6,2.7                           &  0.3,0.7                           &  $M_{\odot}$  \\
$N({\rm H_2})_{\rm 350 \micron}^d$  &  $1.6\times10^{23}$                &  $0.6\times10^{23}$                &  cm$^{-2}$    \\
$N({\rm H_2})_{\rm 3.2 mm}^d$       &  ${\rm 2.0\times10^{25}}$          &  ${\rm 0.9\times10^{25}}$          &  cm$^{-2}$    \\        
$n({\rm H_2})_{\rm 350 \micron}^d$  &  $2.1\times10^6$                   &  $1.0\times10^6$                   &  cm$^{-3}$    \\
$n({\rm H_2})_{\rm 3.2 mm}^d$       &  ${\rm 1.6\times10^{9}}$           &  ${\rm 1.2\times10^{9}}$           &  cm$^{-3}$    \\ 
$\Delta V({\rm N_2H^+})$            &  $1.25\pm0.03$                     &  $0.94\pm0.08$                     &  km s$^{-1}$  \\
$\Delta V({\rm HCO^+})$             &  $1.17\pm0.02$                     &  $1.80\pm0.10$                     &  km s$^{-1}$  \\
$\Delta V({\rm C^{18}O})$           &  $2.6\pm0.1$                       &  $1.4\pm0.2$                       &  km s$^{-1}$  \\
\hline
\end{tabular} \\
$a.${ For ORI8nw\_2, $L_{\rm stars}$ is a summation of YSOs within the 350 $\micron$ emission region. The luminosity of the YSOs (123, 116 and so on) was integrated from their emissions between 3.6 as 8.0 $\micron$ as interpolated from the emissions at four IRAC bands (\citet{galfalk08}, Table 11 therein). And the luminosities of 117 and 120 are directly taken form their Table 6.}  \\
$b.${ From Paper II.} \\
$c.${ In each case, the first and second values are calculated from the line widths of ${\rm N_2H^+}$ and ${\rm C^{18}O}$ respectively.  } \\
$d.${ The values at the core center $(0'',0'')$. Since the emission region is not fully resolved with the CARMA beam size, the derived $N({\rm H_2})$ and $n({\rm H_2})$ may represent lower limits. } \\
\end{minipage}
}
\end{table}

\begin{table}   
{\scriptsize
\centering
\begin{minipage}{150mm}
\caption{Chemical Properties of the ${\rm N_2H^+}$, ${\rm HCO^+}$ and CO.}
\begin{tabular}{lll}
\hline\hline
Species                      &  peak                          &  average$^b$                \\
\hline                                                                                      
\quad(ORI8nw\_2)             &  \quad                         &  \quad                      \\
$N({\rm N_2H^+})$            &  $(1.3-14.0)\times10^{14}$     &  $(1.6-16)\times10^{13}$    \\
$N({\rm HCO^+})$             &  $(3.0-7.3)\times10^{13}$      &  $(0.4-1.3)\times10^{13}$   \\
$N({\rm CO})^a$              &  $-$                           &  $(1.0-2.2)\times10^{18}$   \\
${\rm [CO]}$                 &  $-$                           &  $(0.6-1.3)\times10^{-5}$   \\                                                                  
${\rm [N_2H^+]/[HCO^+]}$     &  $-$                           &  $4-12$                     \\
\hline                                                                                      
\quad(ORI2\_6)               &  \quad                         &  \quad                      \\
$N({\rm N_2H^+})$            &  $(0.68-7.1)\times10^{14}$     &  $(0.8-6.8)\times10^{13}$   \\
$N({\rm HCO^+})^c$           &  $(3.1-6.1)\times10^{12}$      &  $(0.7-1.7)\times10^{12}$   \\
$N({\rm CO})^a$              &  $-$                           &  $(2.2-4.5)\times10^{17}$   \\
${\rm [CO]}$                 &  $-$                           &  $(1.6-5.0)\times10^{-6}$   \\
${\rm [N_2H^+]/[HCO^+]}^c$   &  $-$                           &  $11-40$                    \\
\hline  
\end{tabular}
 \\
Note. For each quantity, the variation range is calculated at two temperature limits, including the currently fitted dust temperature and the bolometric temperatures provided by \citet{manoj13}. The temperature limits are 23-63 K for ORI8nw\_2 and 19-59 K for ORI2\_6. \\
$a.${ For each core, it is calculated from the CSO ${\rm C^{18}O~(2-1)}$ line at the 350 $\micron$ core center at the two temperature limits, and with the assumption that ${\rm [^{12}C]/[^{18}C]=375}$. }  \\
$b.${ For ${\rm N_2H^+}$ and ${\rm HCO^+}$, the value represents an average over an area equal to the CSO beam at 350 $\micron$ $(9'')$. }  \\
$c.${ The ${\rm HCO^+}$ is only marginally detected in ORI2\_6. Therefore its $N({\rm HCO^+})$ may represent an upper limit while ${\rm [N_2H^+]/[HCO^+]}$ as a lower limit.}\\  
\end{minipage}
}
\end{table}


{\small
\begin{figure}
\centering
\includegraphics[angle=0,width=1\textwidth]{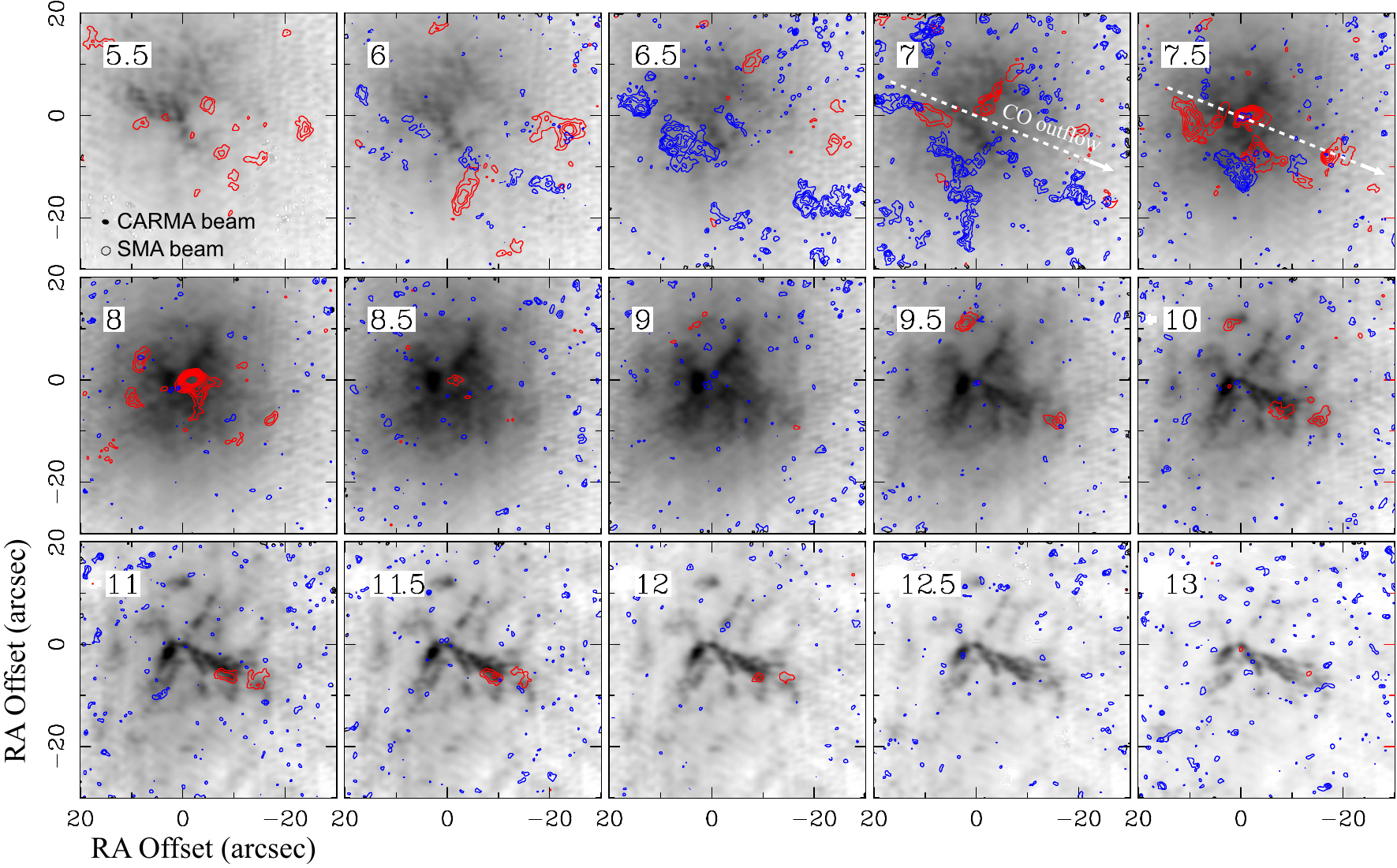} \\
\caption{Channel images of the molecular emissions in ORI8nw\_2. The blue and red contours represent the
${\rm N_2H^+}$ and ${\rm HCO^+}$ and $(1-0)$ and emissions, respectively
(please see the electric version for the colored version). The ${\rm N_2H^+}$
emission is from $F_1=1-1$ component. The contours are 4, 6, 8, 10, 12
$\sigma$. For the ${\rm HCO^+}$, the contours are 4, 8... 28 $\sigma$
($\sigma=0.02$ Jy beam$^{-1}$ for the both lines). The gray scale image is the
CO $(2-1)$ (SW07). The white dashed line indicates the CO outflow direction.
the CARMA beam is in the bottom right and the velocity in km s$^{-1}$ is in
the upper left. The CARMA and SMA beams are plotted in filled and hollow
ellipses respectively, on the bottom of the first panel. For the both sources,
the central position is the 3.2 mm continuum center. The coordinates are
${\rm RA=05^h36^m18\fs79}$, ${\rm Dec=-06\arcdeg22\arcmin10\farcs4}$ for ORI8nw\_2, and
${\rm RA=05^h35^m13\fs43}$, ${\rm Dec=-05\arcdeg57\arcmin57\farcs9}$ for ORI2\_6.} 
\end{figure}

\begin{figure}
\centering
\includegraphics[angle=0,width=0.9\textwidth]{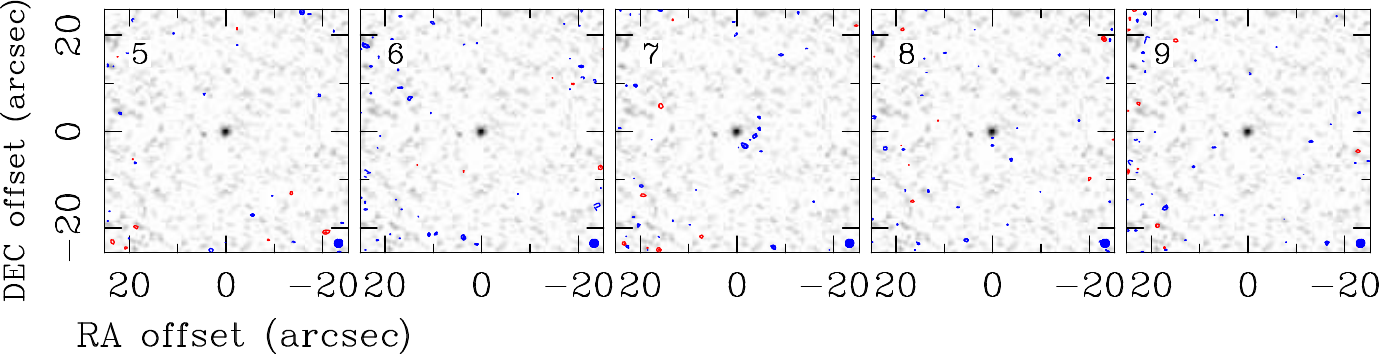} \\
\caption{Same as Figure 1 but for ORI2\_6 except that the CO $(2-1)$ is absent. The background is the 3.2 mm continuum image.}
\end{figure}

\begin{figure}
\centering
\includegraphics[angle=0,width=0.9\textwidth]{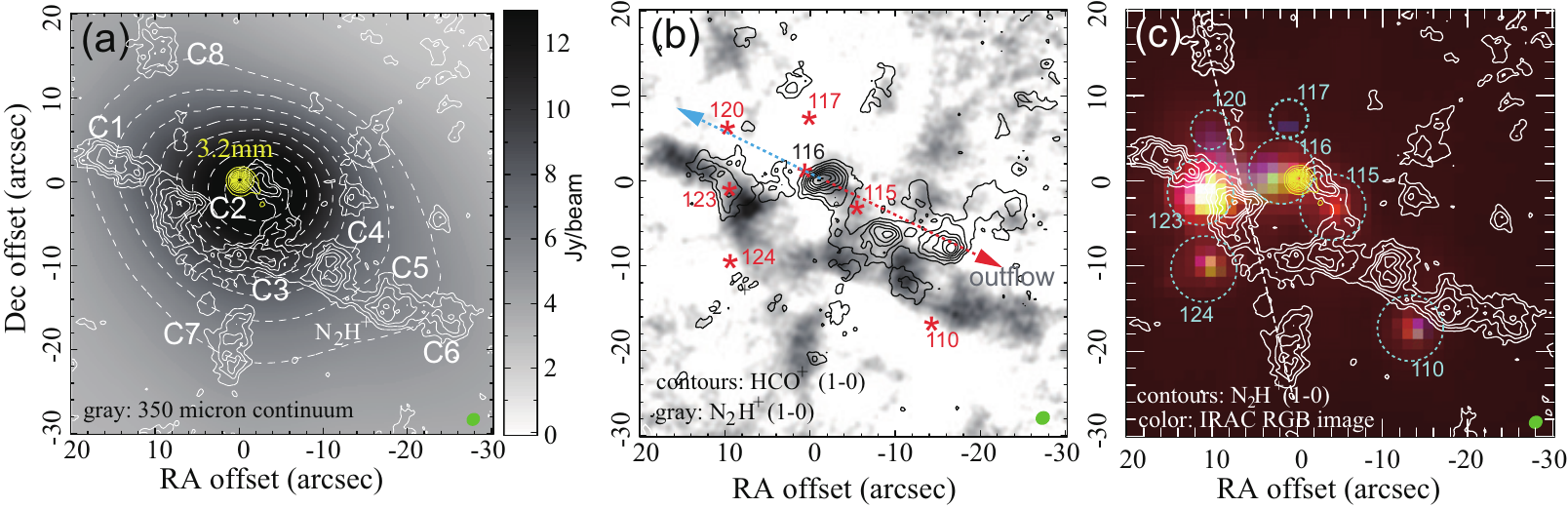} \\
\caption{{\small The velocity-integrated images of the ${\rm N_2H^+}$ and ${\rm HCO^+}$ emissions in ORI8nw\_2.
(a) The ${\rm N_2H^+}$ (thin white contours) and 3.2 mm (thick yellow contours) overlaid on the CSO 350 $\micron$ continuum (gray scale and dashed contours);
(b) The ${\rm HCO^+}~(1-0)$ emission (contours) overlaid on the ${\rm N_2H^+}~(1-0)$ emission (gray). The
dashed arrows indicates the direction of the collimated NE-SW CO outflow \citep{sw07}. The asterisks indicate the positions of the YSOs. (c) ${\rm N_2H^+}$ overlaid on the IRAC RGB image for ORI8nw\_2. For the ${\rm N_2H^+}$ and ${\rm HCO^+}$ emissions, the contours are from 4 $\sigma$ in step of 2 $\sigma$ ($\sigma=0.06$ Jy beam$^{-1}$
km s$^{-1}$ for ${\rm N_2H^+}$ 0.04 Jy beam$^{-1}$ km s$^{-1}$ for ${\rm HCO^+}$). For the 3.2 mm continuum, the contours are 10, 30, 50, 70, 90 \% of the maximum intensity (0.398 Jy beam$^{-1}$). The ${\rm N_2H^+}$ emissions are integrated from $F_1=1-1$ component (see Figure 6 for the spectra). 
component is adopted to plot the ${\rm N_2H^+}$ image. This component has a moderate
optical depth and a high intensity at the same time thus can better exhibit the
gas distribution than the other two components.}}
\end{figure}

\begin{figure}
\centering
\includegraphics[angle=0,width=0.9\textwidth]{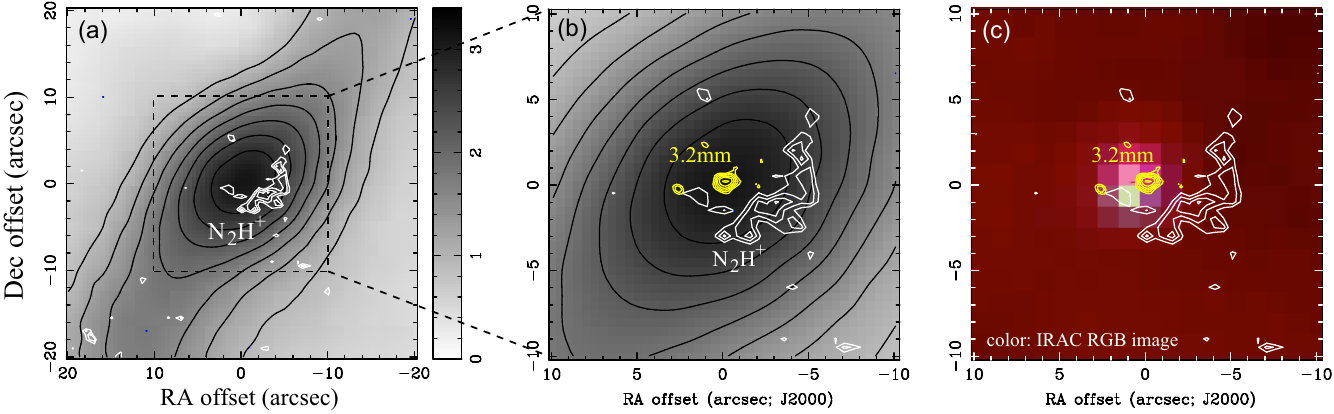} \\
\caption{Same as Figure 3 but for ORI2\_6. The flux density of the 3.2 mm continuum peak is 0.114 Jy beam$^{-1}$.}
\end{figure}

\begin{figure}
\centering
\includegraphics[angle=0,width=0.9\textwidth]{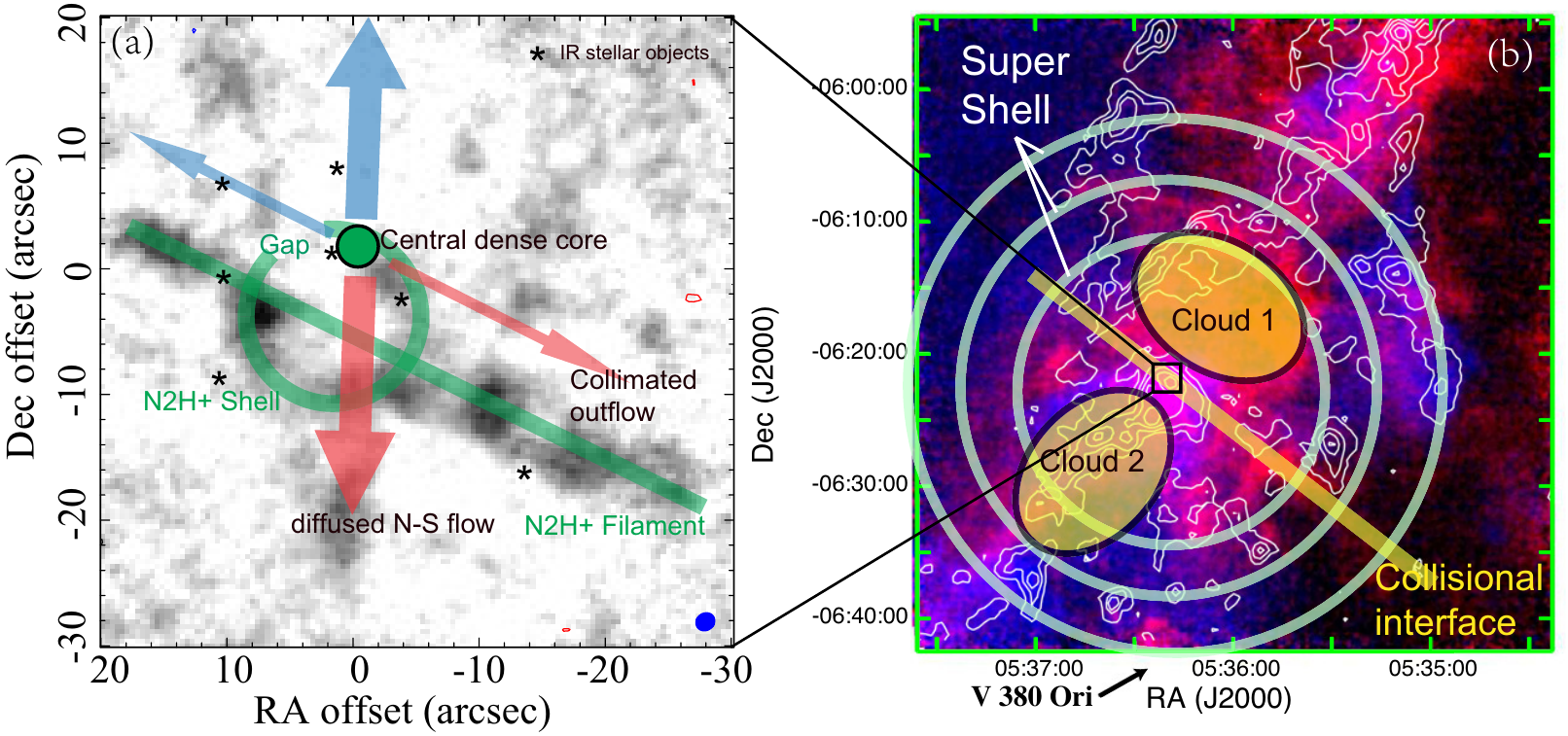} \\
\caption{(a) Illustration for the gas distribution in the ORI8nw\_2 region.
The thick green line and circle shows the distribution and identified structures in ${\rm
N_2H^+}$ $(1-0)$ emission. The blue and red arrows indicates the direction of the CO
outflows observed in SW07. The central dense core and the IR sources are also
labeled in the Figure. (b) The condition of cloud-cloud collision on a larger scale shown in \citet{nakamura12}, with the black box indicating the the region in (a). }  
\end{figure}

\begin{figure}
\centering
\includegraphics[angle=0,width=0.8\textwidth]{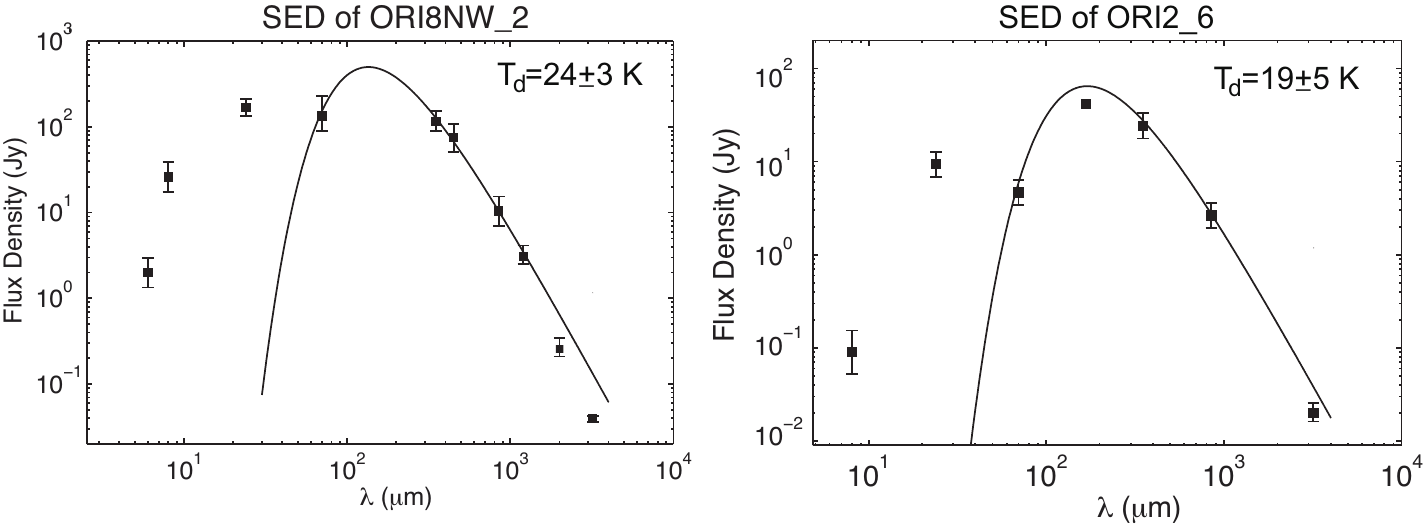} \\
\caption{Spectra energy distribution (SED) of the two cores. The solid line represents the best-fit SED curve For ORI8nw\_2, the is fitted using the observed flux from 70 $\micron$ to 1.3 mm. For ORI2\_6, it is from observed flux at 70 $\micron$, 350 $\micron$, and 850 $\micron$. For the both cores, the IRAC 8 $\micron$, MIPS 24 $\micron$ and CARMA 3.2 mm fluxes are also presented in the figure but not included in the fitting.}  
\end{figure}

\begin{figure}
\centering
\includegraphics[angle=0,width=0.9\textwidth]{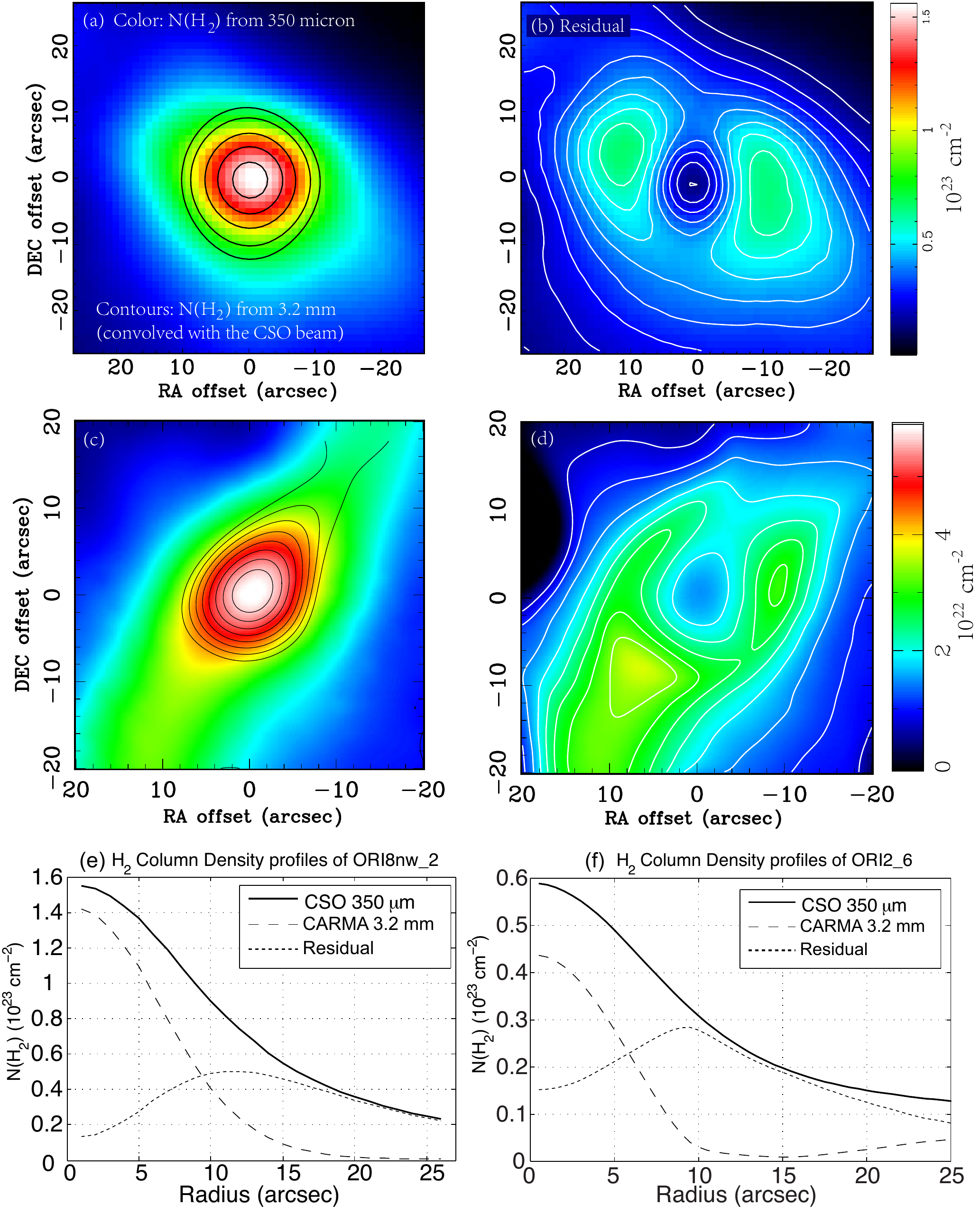} \\
\caption{(a) H$_2$ column density for ORI8nw\_2 calculated from the 350 $\micron$ (false-color
image) and the 3.2 mm emission (contours). The 3.2 mm emission is convolved
with the CSO beam size (9 arcsec); (b) The residual $N({\rm H_2})$
distributions after the component of 3.2 mm emission is subtracted from the 350 $\micron$ component. (c) and (d): same as (a) and (b) but for ORI2\_6; (e) and (f): the H$_2$ column density profiles averaged at
each radius for the two cores.} 
\end{figure}

\begin{figure}
\centering
\includegraphics[angle=0,width=0.8\textwidth]{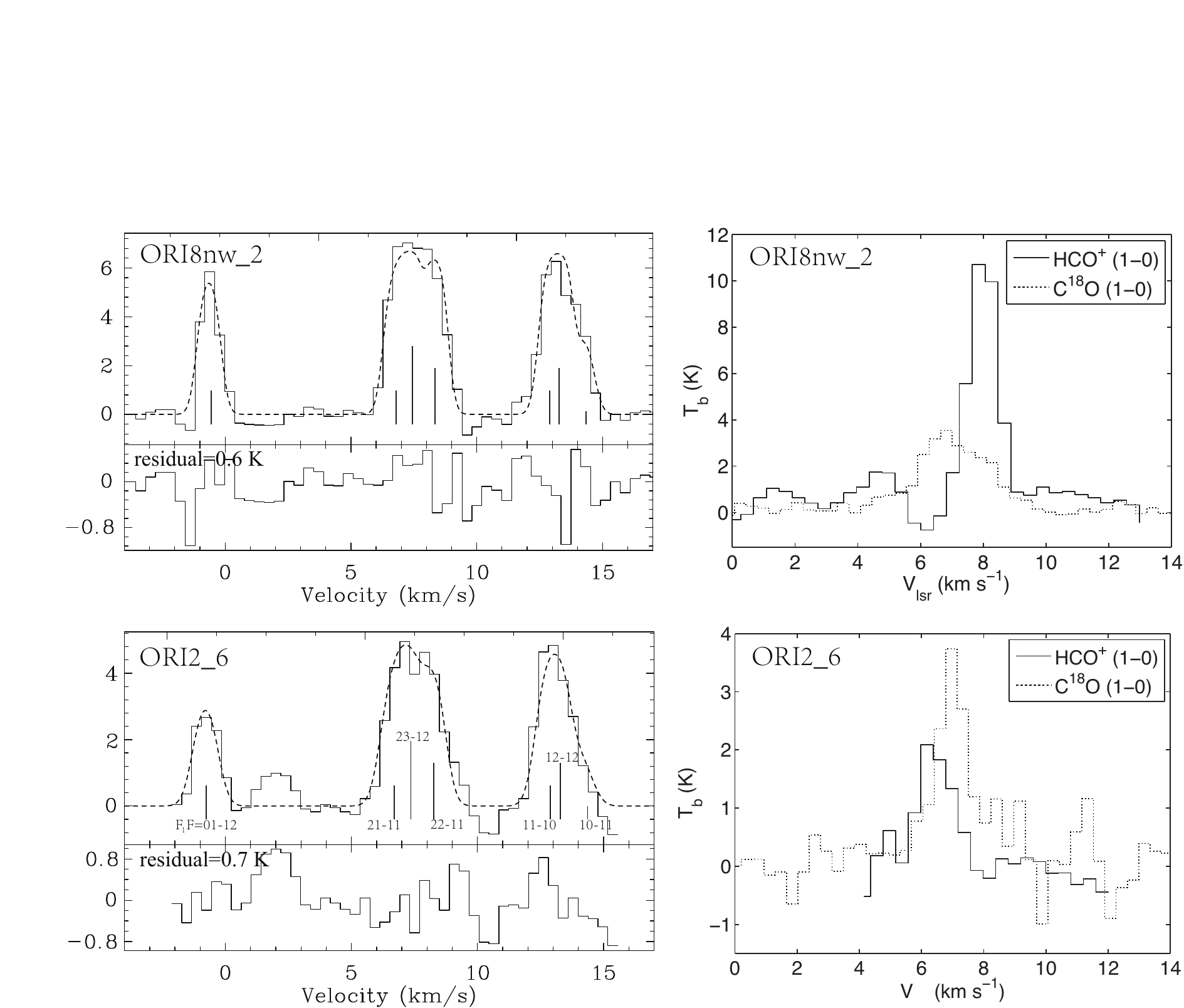} \\
\caption{{\bf Left panels:} the ${\rm N_2H^+}~(1-0)$ line and hyperfine-fitting for ORI8nw\_2 (upper-left panel) and ORI2\_6 (lower-left panel). In each panel, the solid line represents the observed line profile at the emission peak, the dashed line represents the best-fit line profile. The vertical solid lines represents theoretical hyper-fine components, with the vertical length proportional to the relative intensity of each component. In each panel, the residual after fitting is also presented bellow the spectrum. {\bf Right panels:} the ${\rm HCO^+}~(1-0)$ spectrum with the CARMA (solid line) and ${\rm C^{18}O}~(2-1)$ spectrum with the CSO (dashed line, Ren et al. in prep.) towards the continuum peak in the two cores. }  
\end{figure}

\begin{figure}
\centering
\includegraphics[angle=0,width=0.9\textwidth]{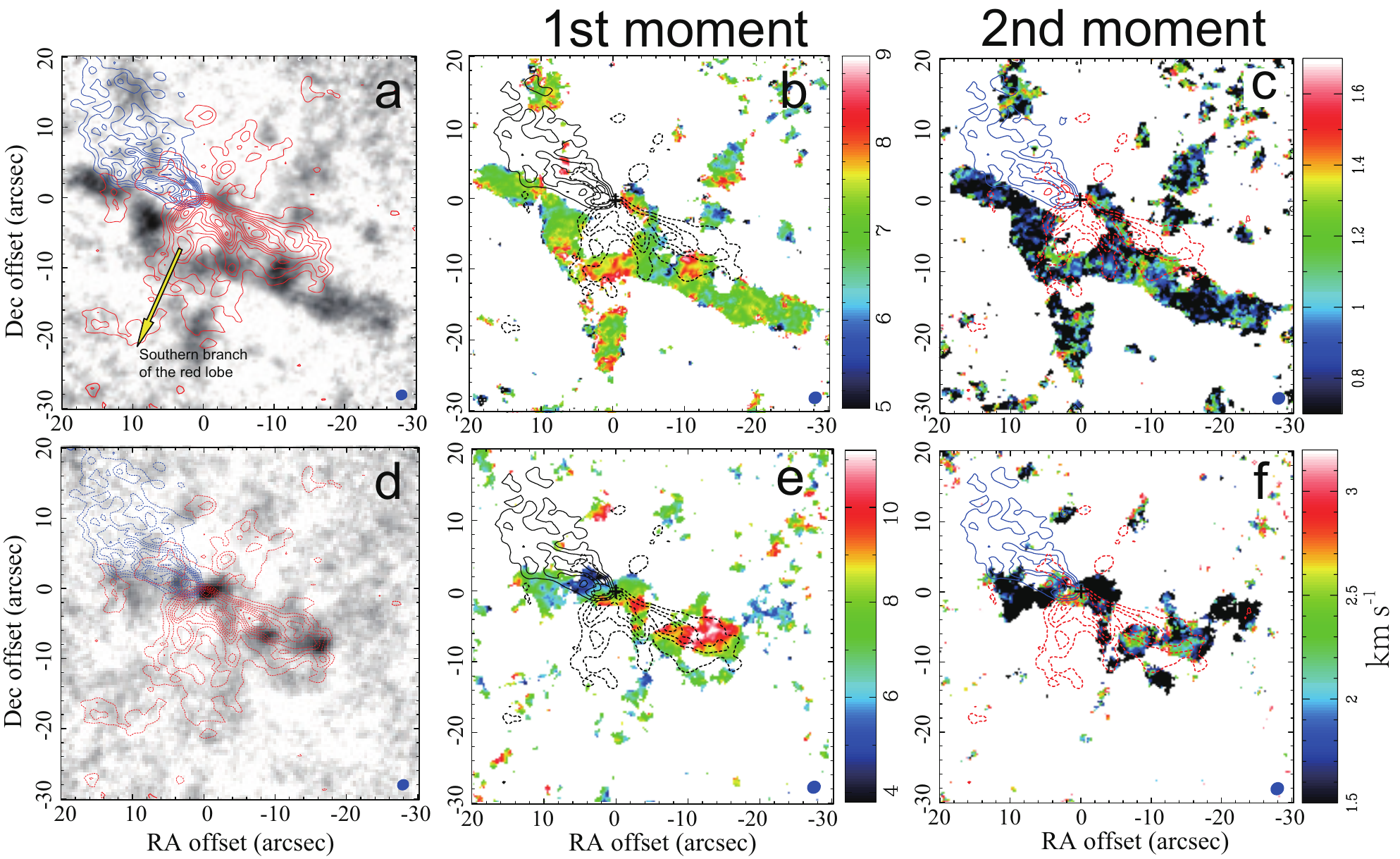} \\
\caption{Outflow and the velocity field of the molecular lines in ORI8nw\_2 (please see the electronic version for the colored images). (a) The CO $(2-1)$ outflow (blue and red contours, SW07) overlaid on the ${\rm N_2H^+}$ emission (gray scale). (b) The CO outflow overlaid on the ${\rm N_2H^+}$ first-moment map (intensity-weighted velocity distribution). (c) The CO outflow overlaid on the
${\rm N_2H^+}$ second-moment map (intensity-weighted velocity dispersion). (d) to (f) are the same as (a) to (c) but for the ${\rm HCO^+}~(1-0)$ emission. }  
\end{figure}

\begin{figure}
\centering
\includegraphics[angle=0,width=0.7\textwidth]{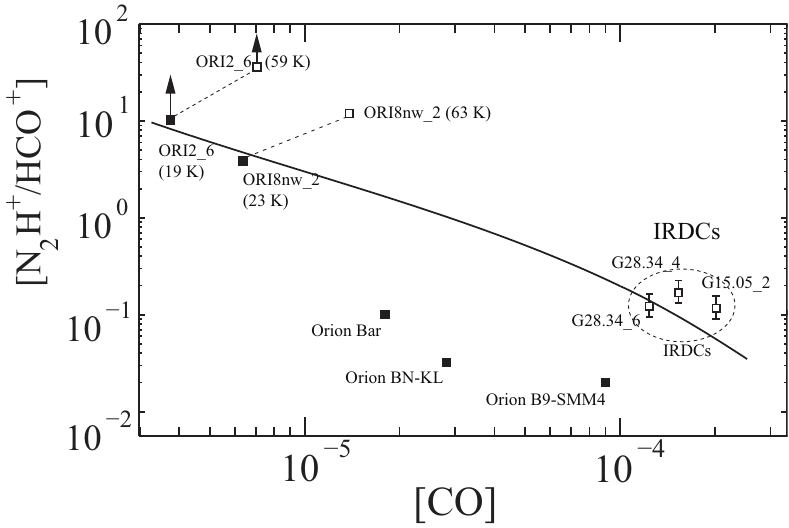} \\
\caption{The model-predicted ${\rm [N_2H^+]/[HCO^+]}$ abundance ratio as a function of the CO abundance \citep{jorgensen04} and the observed values in eight different sources. The modeled curve is plotted in solid line. The data in our two cores and three additional Orion cores \citep{ungerechts97,miettinen12} are plotted in filled squares. For our two cores, the abundances are also calculated at higher temperature values \citep{manoj13} and the results are plotted in open squares. The upper arrow in ORI2\_6 indicates lower limit due to the marginally detected ${\rm HCO^+}$. The data in three IRDCs \citep{vasyunina11,sanhueza12} are plotted with the open squares and encircled with the dashed ellipse.}  
\end{figure}

}

\end{document}